\documentclass[preprint,preprintnumbers,amsmath,amssymb]{revtex4}

\usepackage{graphicx}
\usepackage{dcolumn}
\usepackage{bm}
\begin{document}


\title{Electrodynamics effects in $3+1$ dimensions induced by interactions in Topological Insulators} 

\author{ D.Schmeltzer}

\affiliation{Physics Department, City College of the City University of New York \\
New York, New York 10031}




\begin{abstract}

 We compute  the optical  response of an interacting topological insulator in three space dimensions. The interactions are induced by a chiral charge density wave and the interacting system is invariant under an arbitrary chiral transformation. We show that the chiral phase is gauged away from the action. For the case that the bands are inverted,  the arbitrary phase of the Fujikawa integration measure is fixed by the charge density wave.  We find that   the magnetoelectric  response which is generated by the integration measure breaks  the time reversal symmetry.  At strong  interactions  the  time reversal is restored  and  the magnetoelectric effect  is equivalent to the one obtained in topological insulators in  four space dimensions without interaction.
This effect can be observed  by measuring the Faraday rotation as a function of an external stress. 

\end{abstract}

\pacs{...}


\maketitle
\textbf{I. Introduction}

\vspace{0.2 in}
 
Insulators with spin-orbit interactions can give rise to either regular insulators or to topological insulators ($TI$).
The low energy excitations of the insulators  $Bi_{2}Se_{3}$ and $Bi_{2}Te_{3}$   can be approximated  by a  two-band model with spin-orbit interactions which respect the time reversal symmetry. One finds that when  bands at the $\Gamma$ point are inverted  the system is a  $TI$ according to the classification of the second Chern number \cite{Nakahara} or band invariants introduced in ref.  \cite{Kane}. One of the possible experimental evidence came from the Faraday effect, when light passes from a medium  with  $\theta=0$ (normal insulator) to  a medium with  $\theta\neq 0$ a rotation of the polarization is expected.
Since  time reversal symmetry is not broken the  rotation must  correspond to a topological angle  $\theta=\pm\pi$  \cite{Wilczek,Zhangaxion,Zhangnew}. One of the  experimental  difficulty arises from  the fact that the  experiments are performed in three space dimensions and the second Chern topological invariant exists in four space dimensions.

A formal solution to this problem was given in  \cite{Zhangaxion, Zhangnew,ZhangField}   which suggested  to use dimensional reduction. One performs the computations in four space dimensions and then compactifies one of the coordinates to a small circle.  Other solutions  based on a macroscopic polarization on the surface of the sample have been proposed \cite{ortiz,Vanderbilt}. An adiabatic  approach  to the polarization has been given in \cite{Niu} for the second Chern number. 
The relation between the topological angle $\theta$ and quantization has been discussed  in \cite{Franzaxion}. 
Only recently the effect of interactions on $TI$ has  been considered  \cite{Haijun,Stern}. In particular, the question of charge  density wave instability on the surface of $TI$ has been investigated in ref. \cite{Stern}.
The effect  of an  applied  magnetic field   on the surface of a topological insulator gives rise to  a simple relation between  the   Faraday  and Kerr rotations  \cite{MacDonald}.
 
The main purpose of  this paper is to present a   derivation for  the  optical response (magnetoelectric  effect) of $TI$ in three space dimensions. We obtain this result  using bond interactions   and  chiral  currents   for   insulators with inverted bands. 
The interactions give rise to  an effective action which is controlled  by a bond order parameter. The action of the model   with integration measure is invariant  under an arbitrary chiral transformation  \cite{Fujikawa}.  The saddle point  and the fluctuations of the bond effective action fix the coefficient of the chiral transformation.  As a result one obtains an  electromagnetic  action  which breaks the time reversal symmetry.  The coefficient of the  electromagnetic term is determined by integration (governed by  the  saddle point) which fixes the topological angle   $\theta$.  Experimentally the topological angle is controlled  by the coupling constants and  external  perturbations such as stress \cite{Young}.  The value of $\theta$  determines   the Faraday rotation    between  two regions: the first  region consists of  an interacting $TI$ and  the second  region  represents a non-interacting insulator.  The value of $\theta$ in turn is determined by the integration over the bond order parameter. At particular  values of   $\theta=\pm\pi$, time reversal symmetry is restored and the system is a  $TI$. For $\theta=0$ we have  a regular insulator.

The contents of the paper are as follows: In Sec. II we introduce the model for the $TI$ in the presence of a bond interaction which, with the help of the Hubbard-Stratonovich transformation, is represented as a chiral charge density wave  field $\varphi(\vec{r})$.
Section III represents the central part of this paper, where we include the external electromagnetic field $\vec{A}^{ext}$ in the  action. We perform a chiral transformation $e^{i\lambda(\vec{r})\gamma^{5}}$  for an arbitrary field $ \lambda(\vec{r})$.  Since the partition function is invariant under this transformation, the  action and the integration measure are modified.  The integration measure generates the term  $\lambda(\vec{r})\vec{E} \cdot\vec{ B}$  which breaks the time reversal  symmetry.  The fermionic action $S$ contains a modified bond order parameter which depends on the field  $\lambda(\vec{r})$.  The value of $\lambda(\vec{r})$  can be fixed by demanding that the transformed  bond order parameter vanishes.  As a result, the term  $\lambda(\vec{r})\vec{E} \cdot\vec{ B}$  becomes a function of $\lambda(\varphi(\vec{r}))$. We perform   a saddle point integration over the bond order parameter   and obtain the electromagnetic response function $\theta(\frac{e^2}{2\pi h })\vec{E} \cdot\vec{ B}$ (where $\theta$ is a function of the  fluctuation fields around the saddle point).  Section  IV contains our main conclusions.

\vspace{0.2 in}

\textbf{II. Topological Insulator in three space dimensions }

\vspace{0.2 in}
We will   compute  the   optical    response of  the $TI$ materials  $Bi_{2}Se_{3}$ and $Bi_{2}Te_{3}$. The low energy bands  consist  of  four  projected states,  the conduction and valence  states $|P1^{+}_{-},\pm\frac{1}{2}\rangle$ and  $|P2^{+}_{-},\pm\frac{1}{2}\rangle$  near the Fermi surface at the $\Gamma$ point  \cite{Wei,Chao,XingLiu}. Due to the strong spin-orbit coupling  the level  $|P1^{+}_{-},\pm\frac{1}{2}\rangle$ is pushed down while $|P2^{+}_{-},\pm\frac{1}{2}\rangle$ is pushed up resulting in a band inversion.  Using   the notation  $|orbital=\tau=1\rangle\otimes|spin=\sigma=\uparrow,\downarrow\rangle\equiv |P1^{+}_{-},\pm\frac{1}{2}\rangle$  
and $|orbital=\tau=2\rangle\otimes|spin=\sigma=\uparrow,\downarrow\rangle\equiv |P2^{+}_{-},\pm\frac{1}{2}\rangle$ 
we obtain the effective  Hamiltonian  $\hat{h}^{3d}$ at  the  $\Gamma$ point: 
$\hat{h}^{3d}=k_{2}(\sigma^{1}\otimes \tau^{1})-k_{1}(\sigma^{2}\otimes \tau^{1})+\eta k_{3}(\sigma^{3}\otimes \tau^{1}) + M(\vec{k})(I\otimes \tau^{3})$,  
where $ M(\vec{k})-Bk^2$ determines whether the insulator is trivial or topological.  For $\frac{ M(0)}{B}>0$ we have   a $TI$ (an insulator with an inverted gap)  and a regular insulator for $\frac{ M(0)}{B}<0$ \cite{Kane,Zhangnew}. 
The band anisotropy in the $z$ direction is given by $\eta\ll1$.  Due to Nielsen-Ninomiya theorem  the total number of Dirac points must be even. The other Dirac points are not observed since the Hamiltonian is linearized  around the $\Gamma$ point and contains additional   non-relativistic terms  \cite{Advances}. The eigenvalues in the vecinity of the $\Gamma$ point   are given by $E(\vec{k})=\pm \sqrt{k^2_{1}+k^2_{2}+ \eta^2 k^2_{3}+M(\vec{k})^2}$. We  extend  the model (with a single Dirac point)  to  a torus  and demand the momentum periodicity $-\pi\leq  k_{1}\leq \pi$ ,$-\pi\leq  k_{2}\leq \pi$ ,$-\pi\leq  k_{3}\leq \pi$. 
It is  convenient to perform a unitary transformation   $U=( e^{i\frac{\pi}{4}\sigma^3}\otimes  I)$: 
\begin{equation}
h^{3d}=U^{-1}\hat{h}^{3d}U=  k_{1}\alpha^{1}+k_{2}\alpha^{2}+\eta k_{3}\alpha^{3}+M(\vec{k})\beta , 
\label{ed}
\end{equation}
where the matrices $\alpha$ and $\beta$ are given by: $\alpha^{i}=(\sigma^{i}\otimes\tau_{1})$, $i=1,2,3$ and  $\beta=(I\otimes\tau_{3})$.

Next we include the bond interactions in three space dimensions.
We consider    a particular  type of  bond interaction $H_{int.}$  in three space dimensions which we describe using    the four component   
 spinor $\Psi(\vec{r})=[\Psi_{\tau=1}(\vec{r}),\Psi_{\tau=2}(\vec{r})]^{T}$, where $\Psi_{\tau}(\vec{r})$ is a two component spinor with $\sigma=\uparrow,\downarrow$, $\Psi_{\tau}(\vec{r})=[\Psi_{\tau,\uparrow}(\vec{r}),\Psi _{\tau,\downarrow}(\vec{r})]^{T}$ and $n_{\tau,\sigma}(\vec{r})=\Psi^{\dagger}_{\tau,\sigma}(\vec{r})\Psi_{\tau,\sigma}(\vec{r})$  represents the fermion number for the orbital $\tau$ and spin $\sigma$.
\begin{eqnarray}
&&H_{int}= \frac{-U_{eff}}{2}\int d^3r
\nonumber\\&&\left[\sum_{\sigma=\uparrow,\downarrow} \left(\Psi^{\dagger}_{\tau=1,\sigma}(\vec{r})\Psi_{\tau=2,\sigma}(\vec{r})- \Psi^{\dagger}_{\tau=2,\sigma}(\vec{r})\Psi_{\tau=1,\sigma}(\vec{r})\right)\right]^{\dagger}\left[ \sum_{\sigma=\uparrow,\downarrow}\left(\Psi^{\dagger}_{\tau=1,\sigma}(\vec{r})\Psi_{\tau=2,\sigma}(\vec{r})- \Psi^{\dagger}_{\tau=2,\sigma}(\vec{r})\Psi_{\tau=1,\sigma}(\vec{r})\right)\right]\nonumber\\&&
=
\frac{U_{eff}}{2}\int d^3r\left[\sum_{\sigma=\uparrow,\downarrow} \Psi^{\dagger}_{\tau=1,\sigma}(\vec{r})\Psi_{\tau=2,\sigma}(\vec{r})-\sum_{\sigma=\uparrow,\downarrow} \Psi^{\dagger}_{\tau=2,\sigma}(\vec{r})\Psi_{\tau=1,\sigma}(\vec{r})\right]^2\nonumber\\&&=
\frac{-U_{eff}}{2}\int d^3r[\Psi^{\dagger}(\vec{r})(I\otimes\tau^{2})\Psi^{\dagger}(\vec{r})]^2  , 
\end{eqnarray}
where $U_{eff}>0$.  
The action  for the  three dimensional $TI$ $h^{3d}$ with  the interaction  $H_{int}$  is given by:
\begin{eqnarray}
&&S=\int_{-\infty}^{\infty} dt\int d^3r\Big[\Psi^{\dagger}(\vec{r},t)\Big[i\partial_{t}+i\alpha^{1}\partial_{1}+i\alpha^{2}\partial_{2}+i\eta\alpha^3\partial_{3}\nonumber\\&&-\beta(M(0)-B\sum_{l=1,2}\partial^{2}_{l})\Big]\Psi(\vec{r},t)+ \frac{U_{eff}}{2}(\Psi^{\dagger}(\vec{r})(I\otimes\tau^{2})\Psi^{\dagger}(\vec{r}))^2\Big] .
\end{eqnarray}
Using the Hubbard-Stratonovich transformation we replace the interaction term  by a chiral density wave  field  $\varphi(\vec{r},t)$.
The interaction corresponds  to  a charge density wave  order which acts between the bands (orbitals). Such an interaction can be induced by phonons which couple between the orbitals and can be enhanced by external stress.  Thus, 



\begin{eqnarray}
&&S=\int_{-\infty}^{\infty} dt\int d^3r\Big[\bar{\Psi}(\vec{r},t)\Big[i\gamma^{0}\partial_{t}+i\gamma^{1}\partial_{1}+i\gamma^{2}\partial_{2} +i\eta \gamma^{3}\partial_{3}\nonumber\\&& -(M(0)-B\sum_{l=1,2}\partial^2_{l})+i\gamma^{5}\sqrt{U_{eff}}\varphi(\vec{r},t) \Big]\Psi(\vec{r},t)\Big]-\frac{1}{2}\int_{-\infty}^{\infty} dt\int d^3r \varphi^2(\vec{r},t) . \nonumber\\&&
\end{eqnarray}
In Eq. (4)  we have  replaced $\Psi^{\dagger}(\vec{r},t)\rightarrow \bar{\Psi}(\vec{r},t)=  \Psi^{\dagger}(\vec{r},t) \gamma^{0}$ and introduced the anti-commuting gamma matrices:
\begin{equation}
\gamma^{0}=\beta\equiv (I\otimes\tau^{3}),\hspace{0.1 in} \gamma^{i}=\beta \alpha^i\equiv (\sigma^{i}\otimes i\tau^2), ~~i=1,2,3,\hspace{0.1 in} \gamma^{5}=(I\otimes\tau^{1}) . 
\label{gama}
\end{equation}

From Hubbard-Stratonovich transformation we observe that  the interaction corresponds  to  a charge density wave  order which acts between the bands (orbitals). Such an interaction can be induced by phonons which couple between the orbitals and can be enhanced by an external stress.  The justification  for the chiral coupling $i\gamma^{5}\sqrt{U_{eff}}\varphi(\vec{r},t)$ must take into consideration the Nielsen-Ninomiya theorem about the even number of Dirac modes which are  absent in our Hamiltonian. The second Dirac point is pushed away by the non-relativistic terms. In ref. \cite{Advances} we have introduced  a lattice model which recovers, in the continuum limit, our model.  The model consists of $H_{lattice}=H_{so}+H_{M}+H_{nrel}$, $H_{so}=iz_{so}\sum_{i=1,2,3}\Psi^{\dagger}(\vec{r})(\sigma^{i}\otimes \tau)\Psi(\vec{r}+\vec{a}_{i})+H.C.$ ( $\vec{a}_{i}$ are the Bravais lattice vectors ). $ H_{so}$ has an even number of Dirac points. When the mass term $H_{M}=M(0)\Psi^{\dagger}(\vec{r})(I\otimes \tau_{1})\Psi(\vec{r})$ is included a mass gap is opened. The non-relativistic term $H_{nrel}=-t\sum_{i=1,2,3}\Psi^{\dagger}(\vec{r})(I\otimes \tau)\Psi(\vec{r}+\vec{a}_{i})$ removes  the Dirac points which are not at the $\Gamma$ point. 

 The mass term  arises due to the momentum difference  between the right and  left mover fermions.  
This can be understood  in the following way:   we expand each orbital using the even number of Dirac points   $\vec{K}_{r}$ ,$r=1,2..even$, $\Psi_{\tau=1}(\vec{r})=\sum_{r=1,..., even} \psi_{\tau=1,r}(\vec{r})e^{i\vec\vec{K}_{r}\cdot\vec{r}}$ and   $\Psi_{\tau=2}(\vec{r})=\sum_{r=1,..., even} \psi_{\tau=2,r}(\vec{r})e^{i\vec\vec{K}_{r}\cdot\vec{r}}$.
 The  unit cell of the crystals consists of  five atoms (two $Bi$ and three $Se$).  The  $Bi$ atoms are displaced by  a  vector $\vec{s}< \vec{a}_{i}$ relative  to the  $Se$ atom ($\vec{a}_{i}$ are the Bravais lattice vectors). We have  two equivalent Se atoms and two equivalent Bi atoms and a lattice distortion can remove this degeneracy.  
 Using a projection method we  approximate  the problem to two  spinors  which correspond to the two  Fermi points ($\vec{K}_{r=-}$ and $\vec{K}_{r=+}$):   
  $\Psi(\vec{r})=[\Psi_{\tau=1}(\vec{r}),\Psi_{\tau=2}(\vec{r})]^{T} \approx [ \psi_{\tau=1,r=+}(\vec{r})e^{i\vec\vec{K}_{r=+}\cdot\vec{r}},\psi_{\tau=2,r=-}(\vec{r})e^{i\vec\vec{K}_{r=-}\cdot\vec{r}}]$.  As a result the  mass term  carries a momentum $\vec{Q}=\vec{K}_{r=+}-\vec{K}_{r=-}$. 
The $H_{M}$  Hamiltonian is modified:  
$H_{M}\approx M(\vec{r})\Psi^{\dagger}_{\tau=1}(\vec{r})\Psi_{\tau=2}(\vec{r})+M^{*}(\vec{r})
\Psi^{*}_{\tau=2}(\vec{r})\Psi_{\tau=1}(\vec{r})\approx M(0)\Psi^{\dagger}(\vec{r})(I\otimes \tau_{1})\Psi(\vec{r})+ +i\sqrt{U_{eff}}\varphi(\vec{r}) \Psi^{\dagger}(\vec{r})(I\otimes \tau_{2})\Psi(\vec{r})$,  where $M(0)\equiv M(\vec{r})+M^{*}(\vec{r})$ and 
$\sqrt{U_{eff}}\varphi(\vec{r})\equiv M(\vec{r})-M^{*}(\vec{r})$.

The transformation from the Dirac fields $ \Psi_{\tau=1}(\vec{r})$,$\Psi_{\tau=2}(\vec{r})$ to the Weyl fields    $\Psi_{R}(\vec{r})$,$\Psi_{L}(\vec{r})$   allows  to make contact  with the  one dimensional systems such as organic materials \cite{Schrieffer}:
$\Psi_{\tau=1}(\vec{r})=\frac{1}{\sqrt{2}}[\Psi_{R}(\vec{r})+\Psi_{L}(\vec{r})]$, $\Psi_{\tau=2}(\vec{r})=\frac{1}{\sqrt{2}}[\Psi_{R}(\vec{r})-\Psi_{L}(\vec{r})].$  
 It has been shown that in one dimension such terms give rise to solitons  and the electronic excitations  carry fractional charges \cite{Schrieffer, Rebi, Goldstone} which is not the case for three dimensions.
The interacting model in the chiral form is given by:
\begin{eqnarray}
&&M(0)( \Psi^{\dagger}_{\tau=1}(\vec{r})\Psi_{\tau=1}(\vec{r})-\Psi^{\dagger}_{\tau=2}(\vec{r})\Psi_{\tau=2}(\vec{r}))+i \varphi(\vec{r})(\Psi^{\dagger}_{\tau=1}(\vec{r})\Psi_{\tau=2}(\vec{r})-\Psi^{\dagger}_{\tau=2}(\vec{r})\Psi_{\tau=1}(\vec{r}))\nonumber\\&&=M(0)[ \Psi^{\dagger}_{R}(\vec{r})\Psi_{L}(\vec{r})+\Psi^{\dagger}_{L}(\vec{r})\Psi_{R}(\vec{r})]+i\varphi(\vec{r})[\Psi^{\dagger}_{R}(\vec{r})\Psi_{L}(\vec{r})-\Psi^{\dagger}_{L}(\vec{r})\Psi_{R}(\vec{r})] . 
\end{eqnarray}
   

We observe  that the combination  $M(0)+i\gamma^{5}\sqrt{U_{eff}}\varphi(\vec{r})$ is chiral invariant (in the chiral notation the combination  $M(0)\pm i\sqrt{U_{eff}}\varphi(\vec{r})$  is chiral invariant). When we perform the chiral transformation the combination   $M(0)+i\gamma^{5}\sqrt{U_{eff}}\varphi(\vec{r})$ is transformed   $(M(0)+i\gamma^{5}\sqrt{U_{eff}}\varphi(\vec{r}))\rightarrow  (M'(0)+i\gamma^{5}\sqrt{U_{eff}}\varphi'(\vec{r}))=e^{-i\lambda\gamma^5}[M(0)+i\gamma^{5}\sqrt{U_{eff}}\varphi(\vec{r})]e^{i\lambda\gamma^5}$.

Integration over the fermion field 
$Z=\int [D\varphi][D\bar{\Psi}][D\Psi] e^{i S}=\int [D\varphi] e^{iS_{eff}[\varphi(\vec{r})]}$ generates an effective  action $S_{eff}[\varphi(\vec{r})]$ which shows that   the combination  $ M^{2}+ U_{eff}\varphi^2(\vec{r})$ is invariant.  
(For the one dimensional case this symmetry implies the existence of domain walls and solitons \cite{Goldstone}.)
  The term $i\gamma^{5}\sqrt{U_{eff.}}\varphi(\vec{r},t)$ breaks the time reversal symmetry. The Hamiltonian with the interactions takes the form: $h^{(3d-int)}[\vec{k}, \sqrt{U_{eff.}}\varphi(\vec{r})]= k_{1}\gamma^{1}+k_{2}\gamma^{2}+\eta k_{3}\gamma^{3}+M(\vec{k})+i\gamma^{5}\sqrt{U_{eff.}}\varphi(\vec{r})$. Time  reversal invariance demands that   $T^{-1}h^{(3d-int)}[\vec{k},\sqrt{U_{eff.}}\varphi(\vec{r})]T=h^{(3d-int)}[-\vec{k},\sqrt{U_{eff.}}\varphi(\vec{r})]$  where  $T=i(\sigma_{2}\otimes I) \textbf{K}$ is the time reversal operator. Performing the time reversal transformation  we find that the term   $i\gamma^{5}\sqrt{U_{eff.}}\varphi(\vec{r},t)$  indeed breaks  the time reversal symmetry: $T^{-1}h^{(3d-int)}[\vec{k},\sqrt{U_{eff.}}\varphi(\vec{r})]T=h^{(3d-int)}[-\vec{k},-\sqrt{U_{eff.}}\varphi(\vec{r})]\neq h^{(3d-int)}[-\vec{k},\sqrt{U_{eff.}}\varphi(\vec{r})]$.

 The computation of  the magnetoelectric response is obtained by expanding the action in terms of the external electromagnetic fields. This calculation is done in the following way: we  restrict the chiral transformation such that the transformed coefficient  $i\gamma^{5}$ is {\it absent} in the action. This means that the contribution from the action will not contain terms which violate the time reversal symmetry. The electromagnetic part which violates the time reversal symmetry is due to the path integration measure \cite{Fujikawa}.   The  chiral transformation which eliminates  the  term  $i\gamma^{5}$ from the action, restricts  the  chiral transformation   to  $2\lambda=\pm\pi$.
 Consequently, the  chiral transformation is restricted   to a  $Z_{2}$  transformation.  The  $Z_{2}$ symmetry is  broken  spontaneously,  and we can expand the  effective action around the broken symmetry  state. For this case we have  zero Goldstone modes  but no  solitons.

\vspace{0.2 in}

\textbf{III. Computation of the magnetoelectric  response for a $TI$   in three  space dimensions} 

\vspace{0.2 in}
In this section we will compute the electromagnetic response for the interacting  `inverted' insulator $\frac{ M(0)}{B}>0$  in three space  dimensions.
Since the  interaction  model breaks  the time reversal  symmetry  we obtain a magnetoelectric   response. For  the strong coupling case the proportionality coefficient   is given by $\pm\pi$. As a result we obtain that the magnetoelectric   response does not violate the time reversal symmetry.     We obtain the result using the chiral charge density wave field  $i\gamma^{5} \sqrt{U_{eff}}\varphi(\vec{r})$ and and    chiral transformation $e^{i\gamma^{5} \lambda(\vec{r})}$ with the arbitrary field $\lambda(\vec{r})$. We perform  the    chiral transformation using the existing  results  given in the literature \cite{Fujikawa,Nakahara,Ludwig,Burkov}.  In our problem the saddle point with respect to the field  $i \sqrt{U_{eff}}\varphi(\vec{r})\gamma^{5}$ allows us to fix the value of the  field $\lambda(\vec{r})\equiv  \lambda(\varphi(\vec{r}))$.  Consequently, the saddle point field fixes the coefficient of the magnetoelectric response  $\vec{E}^{ext.}\cdot\vec{B}^{ext}$ and determines  the topological angle $\theta$.
In order to compute the electromagnetic response we will  include  in the action  $S$ the external electromagnetic field,   
$\vec{A}^{ext}(\vec{r},t)=[A^{ext}_{1}(\vec{r}),A^{ext}_{2}(\vec{r}),A^{ext}_{3}(\vec{r},t)]$: 
\begin{eqnarray}
&&S[\bar{\Psi},\Psi,\varphi]=\int_{-\infty}^{\infty} dt\int d^3r\Big[\bar{\Psi}(\vec{r},t)[i\gamma^{0}(\partial_{t}-e A^{ext}_{0})+i\gamma^{1}(\partial_{1}-e A^{ext}_{1})+i\gamma^{2}(\partial_{2}-e A^{ext}_{2})\nonumber\\&& +i\eta \bar\gamma^{3}(\partial_{3}-e A^{ext}_{3})\nonumber\\&& -\left(M(0)-B\sum_{l}\partial^2_{l}\right)+i\gamma^{5}\sqrt{U_{eff}}\varphi(\vec{r},t) ]\Psi(\vec{r},t)\Big]-\frac{1}{2}\int_{-\infty}^{\infty} dt\int d^3r \varphi^2(\vec{r},t) . 
\end{eqnarray}
Next we perform the chiral transformation:
\begin{equation}
\bar{\Psi}(\vec{r},t)\rightarrow  \hat{\bar{\Psi}}(\vec{r},t)= e^{i\lambda(\vec{r},t)\gamma^{5}}\bar{\Psi}(\vec{r},t),\hspace{0.1 in}\Psi(\vec{r},t)\rightarrow \hat{\Psi}(\vec{r},t)= e^{i\lambda(\vec{r},t)\gamma^{5}}\Psi(\vec{r},t)  
\label{chiral}
\end{equation}
The action transforms like:
\begin{eqnarray}
&&S[\bar{\Psi},\Psi,\varphi;M(0)]\rightarrow S[\hat{\bar{\Psi}},\hat{\Psi},\varphi;M(0]\nonumber\\&&=S[\bar{\Psi},\Psi,\varphi;M(0)]+\int dt\int d^3\Big[ \lambda(\vec{r})[\partial_{1}(\bar{\Psi}(\vec{r},t)\gamma^{1}\gamma^5\Psi(\vec{r},t)+\partial_{2}(\bar{\Psi}(\vec{r},t)\gamma^{2}\gamma^5\Psi(\vec{r},t))\nonumber\\&&+\eta\partial_{3}(\bar{\Psi}(\vec{r},t)\gamma^{3}\gamma^5\Psi(\vec{r},t))]
-[M(0)(\cos(2\lambda(\vec{r},t))-1)+\sqrt{U_{eff}}\varphi(\vec{r},t)
 \sin(2\lambda(\vec{r})]\bar{\Psi}(\vec{r},t)\Psi(\vec{r},t)\nonumber\\&&+i[ \sqrt{U_{eff}}\varphi(\vec{r},t) (\cos(2\lambda(\vec{r},t)-1) -M(0)\sin(2\lambda(\vec{r},t))] \bar{\Psi}(\vec{r},t)\gamma^{5}\Psi(\vec{r},t)  +B\bar{\Psi}(\vec{r},t)\gamma^{5}\sum_{l}\partial^2_{l}\Psi(\vec{r},t)   \Big], \nonumber\\&&
\end{eqnarray}
and the partition function obeys the equality:
\begin{equation}
Z=\int [D\varphi][D\bar{\Psi}][D\Psi] e^{i S[\bar{\Psi},\Psi,\varphi]}=\int [D\varphi] [D\hat{\bar{\Psi}}][D\hat{\Psi}] e^{i S[\hat{\bar{\Psi}},\hat{\Psi},\varphi]} .
\label{ze}
\end{equation}
Using the fact that the measure $[D\bar{\Psi}][D\Psi]$ is not invariant under the chiral transformation \cite{Fujikawa}
we obtain the   result in units where $h=c=1$:
\begin{equation}
[D\hat{\bar{\Psi}}][D\hat{\Psi}]=[D\bar{\Psi}][D\Psi]e^{i\frac{e^2}{2\pi }\int_{-\infty}^{\infty} dt\int d^3 r[\lambda(\vec{r})(\vec{E}^{ext}(\vec{r})\cdot\vec{B}^{ext}(\vec{r},t))]} .
\end{equation}
Since the field is arbitrary we can choose this field such that the coefficient of $\gamma^5$ vanishes. As a result the action obtained is time reversal invariant and the  leading electromagnetic contribution is due only  to the measure.
The term  $\gamma^5 \sqrt{U_{eff}} \varphi(\vec{r}) $ becomes    $\gamma^5[ \sqrt{U_{eff}} \varphi(\vec{r},t)\cos(2\lambda(\vec{r},t)) -M(0)\sin(2\lambda(\vec{r},)] $ for a particular choice of $\lambda(\vec{r})$  and the coefficient of   $\gamma^5$   vanishes, 
\begin{equation}
\tan(2\lambda(\vec{r},t))=\frac{ \sqrt{U_{eff}} \varphi(\vec{r},t)}{M(0)} .
\label{equation}
\end{equation}
For  this value of $\lambda(\vec{r},t)$ we  can absorb the interaction field $\varphi(\vec{r},t)$ into a new mass  $\hat{M}(\varphi^2(\vec{r},t))$ (the term which describes the inverted bands): 
\begin{equation}
M(0)\rightarrow M(0)\sqrt{1+\frac{ U_{eff} \varphi^{2}}{M^2(0)}}\equiv M(\varphi^2(\vec{r},t)) . 
\label{mas} 
\end{equation}
The partition function can be written as:
\begin{eqnarray}
&&Z=\int [D\varphi][D\bar{\Psi}][D\Psi]e^{i S[\bar{\Psi},\Psi,M(\varphi^2(\vec{r},t))),0]} e^{i\int_{-\infty}^{\infty} dt\int d^3r(-\frac{\sqrt{U_{eff.}}}{(M(0))}\vec{\partial}\varphi(\vec{r},t)\cdot [\bar{\Psi}(\vec{r},t)\vec{\gamma}\gamma^5\Psi(\vec{r},t)]} \nonumber\\&&
\times e^{i\frac{e^2}{2\pi }\int dt\int d^3 r\Big[\frac{1}{2}ArcTan[\frac{\sqrt{U_{eff.}}}{(M(0))}\varphi(\vec{r},t)]\vec{E}^{ext}(\vec{r})\cdot\vec{B}^{ext}(\vec{r})\Big]}
e^{i\int_{-\infty}^{\infty} dt\int d^3r\frac{- \varphi^2(\vec{r},t)}{2}} , 
\end{eqnarray}
where
\begin{eqnarray}
&&S[\bar{\Psi},\Psi,M(\varphi^2(\vec{r}))),0]=\nonumber\\&&
\int_{-\infty}^{\infty} dt\int d^3r\Big[\bar{\Psi}(\vec{r},t)[i\gamma^{0}(\partial_{t}+ie A^{ext}_{0})+i\hbar\gamma^{1}(\partial_{1}+ie A^{ext}_{1})+i\gamma^{2}(\partial_{2}+ie A^{ext}_{2}) \nonumber\\&&+i\eta \gamma^{3}(\partial_{3}+ie A^{ext}_{3}) -(M(\varphi^2(\vec{r},t)) -B\sum_{l}\partial^2_{l}) +  i\gamma^{5}(0)
]\Psi(\vec{r},t))\Big]\nonumber\\&&\equiv \int_{-\infty}^{\infty} dt\int d^3r\Big[\bar{\Psi}(\vec{r},t)[i\gamma^{0}(\partial_{t}+ie A^{ext}_{0})-h^{3D-eff.}(\vec{k},M(\varphi^2(\vec{r},t)))]\Psi(\vec{r},t))\Big] . 
\end{eqnarray}
 The action  $S[\bar{\Psi},\Psi,M(\varphi^2(\vec{r},t)),0]$ does not break the time reversal symmetry, therefore the integration of the fermions generates only regular Maxwell terms. 
We perform the integration with respect to the fermions and compute the effective potential for zero electromagnetic fields.
\begin{eqnarray}
&&\int [D\bar{\Psi}][D\Psi]e^{i S[\bar{\Psi},\Psi,M(\varphi^{2}(\vec{r},t)),0]}e^{i\int_{-\infty}^{\infty} dt\int d^3r\frac{- \varphi^{2}(\vec{r},t)}{2}}=\nonumber\\&&
e^{-i\int_{-\frac{T}{2}}^{\frac{T}{2}} dt\int d^3r[\frac{\varphi^{2}(\vec{r},t)}{2}+i Ln Det\Big[i\gamma^{0}\partial_{t}-h^{3D-eff.}(\vec{k},M(\varphi^{2}(\vec{r},t))]\Big]}\approx e^{-i\int_{-\frac{T}{2}}^{\frac{T}{2}}\int d^3r\Big[ W^{(3D-eff.)}[\varphi(\vec{r})]\Big]}\nonumber\\&&%
W^{(3D-eff.)}[\varphi(\vec{r})]=\frac{\varphi^{2}}{2}+i\int_{-\infty}^{\infty} \frac{d\omega}{\pi}\int_{0}^{\Lambda} \frac{d^3k}{(2\pi)^3}Ln\Big[\omega^2-[k^2_{1}+ k^2_{2}+\eta^2 k^2_{3}+(M(\varphi^{2},t)-Bk^2)^{2}]+i\epsilon\Big] .\nonumber\\&&
\end{eqnarray}
We introduce the ultraviolet cut-off $\Lambda$  \cite{Shankar} and replace  the effective gap $M(0)\equiv\Lambda\hat{M}(0)$ and the coupling constant $\frac{1}{U_{eff}}\equiv \hat{g}_{eff.}\Lambda^{d-1}$. The field $\varphi(\vec{r},t)$ is replaced with   $\hat{\Phi}(\vec{r},t)$, defined through the relation:
$\sqrt{U_{eff.}}\varphi(\vec{r},t)=\Phi(\vec{r})\equiv (\frac{\Phi(\vec{r},t)}{M(0)})M(0)$,
$\frac{\Phi(\vec{r},t)}{M(0)}\equiv \hat{\Phi}(\vec{r},t)$. 

The effective potential in Eq.~$(16)$ is a function of the sign of the parameter $B$.  The number of zero eigenvalues in the energy spectrum  $E(\vec{k})=\pm\sqrt{[k^2_{1}+ k^2_{2}+\eta^2 k^2_{3}+(\hat{M}(\varphi^{2})-Bk^2)^{2}]}$ determines if the system is a regular insulator or a $TI$. For a  positive mass   $M(0)$  and  $B<0$  the eigenvalues have no zero energy states. Computing  the effective potential for this case ($B<0$) shows that $W^{(3D-eff)}[\varphi(\vec{r})]$ has only a single minimum for  $\varphi(\vec{r})=0$. Therefore the magnetoelectric effect is absent and the system is a regular insulator in agreement with the literature.  For the case $B>0$ ,  $E(\vec{k})$  has zero energy states. The effective potential  $W^{(3D-eff)}[\varphi(\vec{r})]$  for this case has non-zero saddle points giving rise to a magnetoelectric effect and the system is a $TI$  in agreement with the condition  $\frac{M(0)}{B}>0$. 

For the remaining part we restrict our calculation to the $TI$ case  $\frac{ M(0)}{B}>0$.
We compute the  effective potential  for $d=3$ , $0<B\Lambda^2\ll1$  and $\eta\rightarrow 0$:
\begin{eqnarray}
&&W^{(3d-eff)}[\hat{\Phi}(\vec{r})]=\frac{1}{2}\hat{g}_{eff.}(\hat{M}(0))\Lambda^{4}\hat{\Phi}^2(\vec{r})-\frac{2}{(2\pi)^3}\int_{0}^{\Lambda} d^{2}k_{\bot}\int_{-\Lambda}^{\Lambda} dk_{3}\nonumber\\&&\Big[\sqrt{k^2_{1}+k^2_{2}+\eta^2 k^2_{3}+\Lambda^2\hat{M}^{2}(0)(1+\hat{\Phi}^2(\vec{r}))}-\sqrt{k^2_{1}+k^2_{2}+\eta^2 k^2_{3}}\Big]\nonumber\\&&
\frac{W^{(3d-eff)}[\hat{\Phi}(\vec{r})]}{\Lambda^{4}}\approx \frac{1}{2}\hat{g}_{eff.}\hat{M}^{2}(0) \hat{\Phi}^2(\vec{r})-\frac{1}{6\pi^2}\Big[[1+\hat{M}^2(0)(1+\hat{\Phi}^2(\vec{r}))]^{\frac{3}{2}}\nonumber\\&&-[1+[\hat{M}^2(0)(1+\hat{\Phi}^2(\vec{r}))]^{\frac{3}{2}}\Big] . 
\end{eqnarray}
In Fig. 1 we plot the effective potential $\frac{W^{(3d-eff)}[\hat{\Phi}(\vec{r},t)]}{\Lambda^{4}}$ for the  inverted  gap   $\frac{ M(0)}{B}>0$.  
Varying the values of the coupling constant $U_{eff}$  (controlled by the pressure) $\varphi$ decreases below $\pi$. The value    $\varphi<\pi$   corresponds to   a phase with a  broken time reversal symmetry.  At  strong coupling the order parameter takes the values    $\varphi= \varphi^{*}=\pm\pi$   and the  time reversal symmetry is restored.

Using the effective potential $W^{(3D-eff)}[\varphi(\vec{r},t)]=W^{(3D-eff)}[\hat{\Phi}(\vec{r},t)]$ we compute the electromagnetic response:
\begin{eqnarray}
&&Z=\int [D\varphi]e^{-iW^{(3D-eff)}[\varphi(\vec{r},t)]}e^{i\frac{e^2}{2\pi }\int dt\int d^3 r[\frac{1}{2}ArcTan[\frac{\sqrt{U_{eff}}}{(M(0)}\varphi(\vec{r},t)](\vec{E}^{ext}(\vec{r})\cdot\vec{B}^{ext}(\vec{r}))]}\nonumber\\&&\approx e^{i\theta(\frac{e^2}{2\pi })\int_{-\infty}^{\infty} dt\int d^3 r\vec{E}^{ext}(\vec{r})\cdot\vec{B}^{ext}(\vec{r})}\equiv e^{i\delta S^{(3)}_{eff}}.  
\end{eqnarray}
The results in the last equation have been obtained with the help of the saddle point integration.
The effective potential  $W^{3D-eff}[\varphi(\vec{r},t)]$ has a saddle point  $\pm\sqrt{U_{eff}}\varphi\equiv \Phi \neq0$.  
We  perform the integration of the field  $\varphi(\vec{r})$ expanding the action around  the saddle point   $\Phi \neq0$,   $\Phi(\vec{r})=\Phi +\delta \Phi(\vec{r},t)$. The  fluctuation field      $\delta \Phi(\vec{r},t)$ corresponds to  the  zero frequency Goldstone mode. The  Goldstone mode can create a domain wall for the case when the interactions are  restricted to half space. 

In summary, we find that when $\sqrt{U_{eff}}\langle\varphi(\vec{r},t)\rangle=\Phi\neq 0$ the time reversal symmetry is broken and the effective action is given by Eq. $(18)$.  In the limit of strong interaction the expectation value can attain the value  $\sqrt{U_{eff}}\langle\varphi(\vec{r})\rangle=\Phi^{*}=\pm\pi$. At these values of  $\Phi^{*}=\theta=\pm\pi$ and the time reversal symmetry is restored.

In the last part of this paper we want to comment about the possibility of solitons for  three dimensional crystals.  This can be achieved  by replacing the scalar field  $\varphi(\vec{r})$ by a vector   field $\vec{\varphi} (\vec{r})\cdot\vec{S}$ which couples  through the Pauli matrix   $\vec{S}$ to   two  Fermi points  like in  graphene ,  $\hat{\Omega}(\vec{r},t)=[\bar{\Psi}^{F1}(\vec{r},t),\bar{\Psi}^{F2}(\vec{r},t)]$ where $F1$ and $F2$ are the two Fermi points.  
As a result the effective action is modified,
\begin{eqnarray}
&&\hat{S}=\int_{-\infty}^{\infty} dt\int d^3r\Big[\hat{\Omega}(\vec{r},t)[i\gamma^{0}\partial_{t}+i\gamma^{1}\partial_{1}+i\gamma^{2}\partial_{2} +i\eta \gamma^{3}\partial_{3}\nonumber\\&& -(M(0)-B\sum_{l=1,2}\partial^2_{l})+i\gamma^{5}\sqrt{U_{eff}}\vec{\varphi}(\vec{r},t)\cdot\vec{S} ]\Omega(\vec{r},t)\Big]-\frac{1}{2}\int_{-\infty}^{\infty} dt\int d^3r \vec{\varphi}^2(\vec{r},t)\nonumber\\&&
\end{eqnarray}
According to \cite{Goldstone} this action can support monopole  fields which are the analog of stable kinks in higher  dimensions. In the condensed matter the monopole fields are similar to dislocations \cite{Schmeltzer}.

Next we  consider the experimental realization.
For practical applications we will assume that the interaction term  $ \sqrt{U_{eff}}\varphi(\vec{r},t)\gamma^{5}$  is restricted to half space. Such a restriction will always guarantee that the topological angle is not zero.
In order to confirm the topological behavior we have to measure the  nonlinear response given by  the Faraday and    Kerr rotations. The electrodynamics  in the presence of the new term  $\theta \vec{E}\cdot\vec{B}$ has been studied in \cite{Hehl,Karch}. Using this theory one can compute the relation between the Faraday rotation (transmission) and  Kerr rotation (reflection)  that  is relevant to  the topological angle $\theta$.
We predict that the Faraday rotation  together with the Kerr rotation will be sensitive to external perturbations such as pressure. 
Therefore, by changing  the  pressure we can induce a change in the  rotation  and identify the topological angle $\theta=\pm\pi$. According to the  sensitivity reported  (10 $nrad$) in ref. \cite{Kapitulnick}  the  experiments suggested there seem to be achievable. The authors  in ref .\cite{Kapitulnick} use a reflection mirror   to distinguish between a Faraday rotation  which breaks the time reversal symmetry and the one which does not.  (For a system which breaks the time reversal symmetry one finds that due to the reflection the Faraday rotation is twice the original one, and for a system which respects the time reversal symmetry  the angle of rotation after the reflection is zero.)

According to   refs. \cite{Senthil,Nagaosa} the magnetoelectric effect can give rise to  fractional statistics  or magnetic texture  on the surface of the $TI$. 
Therefore   transport measurements on the surface of the $TI$   can differentiate between the different predictions which correspond to different  topological angles. 
\vspace{1.0 in}

\textbf{IV. Conclusions}

\vspace{0.2 in}
We have computed the magnetoelectric effect in the presence of interactions  for a $TI$ in three space dimensions.  We find that for a  particular type of  bond interaction which is described as a chiral charge density wave  we  obtain  an electromagnetic response  which  breaks  the time reversal symmetry.  We find the magnetoelectric  response $\theta(\frac{e^2}{2\pi })\int_{-\infty}^{\infty} dt\int d^3 r[\vec{E}^{ext}(\vec{r})\cdot\vec{B}^{ext}(\vec{r})]$ which depends on the topological angle  $\theta$.  At strong couplings $\theta=\pm\pi$,  the time reversal symmetry is restored and  we obtain that the  electromagnetic response of the three dimensional  $TI$   with interactions is equivalent  to the  non-interacting $TI$ in four space  dimensions.  The saddle point is sensitive  to an external $stress$, therefore this theory can be tested  by  measuring the Faraday  and Kerr rotations of a $TI$ crystal under variable hydrostatic stress.



\clearpage
\begin{figure}
\begin{center}
\includegraphics[width=7.2 in ]{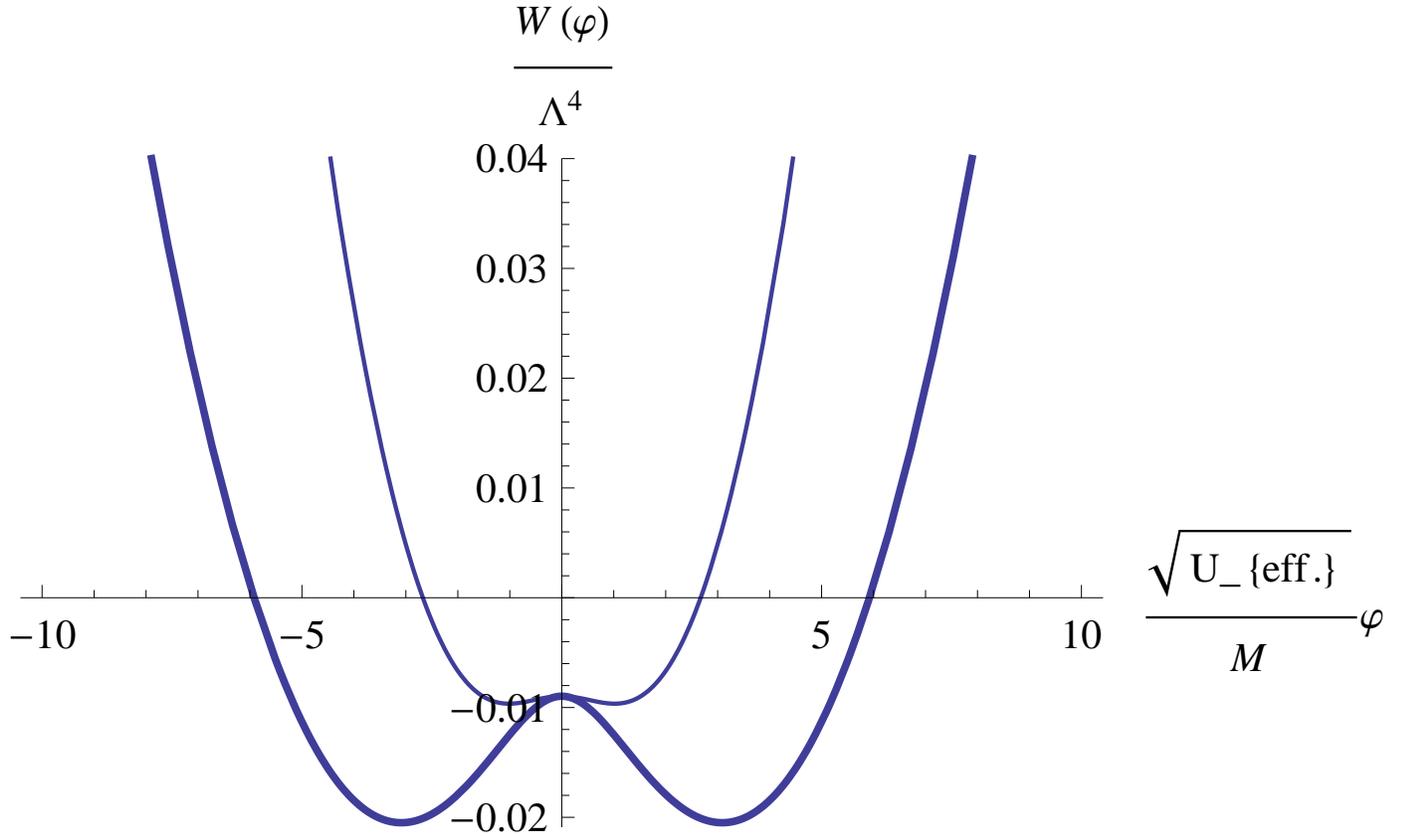}
\end{center}
\caption{The effective  potential $W(\sqrt{U_{eff}}\varphi)$ for $(\hat{M}(0))=0.75$. The lower  graph (the thicker line) represents the effective potential for $\hat{g}_{eff}\equiv\frac{1}{\hat{U}_{eff}}=0.01$. For this case  the saddle point   corresponds to $\sqrt{U_{eff}}\varphi=\varphi^{*}=\pm\pi$. The upper graph  represents the effective potential for a weaker  coupling    $\hat{g}_{eff}\equiv\frac{1}{\hat{U}_{eff}}=0.02$. As a result we find that  $\sqrt{U_{eff}}\varphi \approx\pm 1$.} 
\end{figure}

\end{document}